\documentclass[twocolumn,prb,spanish]{revtex4}

\usepackage{graphicx}
\usepackage{dcolumn}
\usepackage{bm}
\usepackage[spanish]{babel}
\begin{document}

\title{Espectro de energ{\'\i}as, densidad de niveles, polarizaci{\'o}n
 del esp{\'\i}n, propiedades de transporte y {\'o}pticas en puntos 
 cu{\'a}nticos y trampas de {\'a}tomos}
\author{Augusto Gonz{\'a}lez}
\affiliation{Instituto de Cibern{\'e}tica, Matem{\'a}tica y 
F{\'\i}sica  Calle E 309, Vedado, Habana 4, Cuba}
\email{agonzale@cidet.icmf.inf.cu}

\begin{abstract}

Se resume un conjunto de resultados de perfil te{\'o}rico 
[1-10], relativos a 
espectros de energ{\'\i}as, densidad de estados excitados, 
polarizaci{\'o}n del esp{\'\i}n, transporte a trav{\'e}s de barreras de 
potencial, y propiedades {\'o}pticas (absorci{\'o}n en el infra-rojo, 
luminiscencia) de sistemas tales como puntos
cu{\'a}nticos semiconductores y trampas de {\'a}tomos.
La caracter{\'\i}stica fundamental de los sistemas investigados consiste 
en que contienen un n{\'u}mero de part{\'\i}culas entre 4 y 400, lo cual 
hace muy dif{\'\i}cil el c{\'a}lculo de sus propiedades.  En los trabajos
presentados se resalta y se hace amplio uso de la analog{\'\i}a entre estos 
sistemas y los n{\'u}cleos at{\'o}micos por lo que se adaptaron m{\'e}todos  
de la F{\'\i}sica Nuclear tales como: Hartree-Fock
y la RPA para sistemas finitos, el m{\'e}todo BCS y la proyecci{\'o}n de 
Lipkin-Nogami y la 
ecuaci{\'o}n de Bethe-Goldstone. Adem{\'a}s, se resolvi{\'o} la 
ecuaci{\'o}n de Schrodinger utilizando bases de funciones de hasta 
40,000 elementos utilizando el algoritmo de Lanczos y se implementaron 
otros m{\'e}todos como el de Monte Carlo variacional y los aproximantes 
dobles de Pad{\'e}.  Por {\'u}ltimo, se propuso un m{\'e}todo 
estoc{\'a}stico para proyectar la funci{\'o}n de onda BCS.

\vspace{.5cm}
A set of theoretical results [1-10] is reviewed, which concern
calculations of energy spectra, density of energy levels,
spin polarization, transport and optical properties (infrared absorption,
luminescence) of semiconductor quantum dots and atomic traps.
The studied systems contain a number of particles between 4 and
400, thus the calculation of their physical properties is a hard task.
The analogy between these systems and the atomic nuclei is stressed
and used throughout the paper. Common Nuclear Physics methods like Hartree-Fock
and RPA schemes for finite systems, the BCS approach and the 
Lipkin-Nogami projection, and the Bethe-Goldstone equation were
adapted to the present context. On the other hand, the Schrodinger
equation was solved in basis with up to 40,000 functions by means
of the Lanczos algorithm, and other methods like variational
Monte Carlo estimations and two-point Pad{\'e} approximants
were also applied. Lastly, a stochastic projection of the BCS
wave function was implemented.
 
\end{abstract}
\maketitle

\section{Energia de bosones en una trampa}

En el trabajo [1] se calcul{\'o} la energ{\'\i}a del estado base de 
sistemas de hasta 210 bosones ({\'a}tomos con esp{\'\i}n entero) 
confinados en una trampa. El c{\'a}lculo se hizo por dos v{\'\i}as: 
anal{\'\i}ticamente utilizando los denominados
aproximantes dobles de Pad{\'e} [11] y a partir del denominado m{\'e}todo 
variacional de Monte Carlo [12]. 

La novedad del trabajo consisti{\'o} en la extensi{\'o}n del tratamiento 
anal{\'\i}tico empleado en [11] para calcular el espectro de energ{\'\i}as 
del excit{\'o}n (un sistemas de dos part{\'\i}culas: un electr{\'o}n y un 
hueco) a sistemas de cientos de part{\'\i}culas. Para ello se hizo 
necesario calcular series para la energ{\'\i}a en el l{\'\i}mite de 
interacci{\'o}n d{\'e}bil (por teor{\'\i}a de perturbaciones hasta 
segundo orden) y en el l{\'\i}mite de interacci{\'o}n fuerte (oscilaciones 
de una denominada mol{\'e}cula de Wigner). Estas series son luego 
conectadas por medio de aproximantes dobles de Pad{\'e}.

Por otro lado, el m{\'e}todo variacional de Monte Carlo combina el
principio variacional de Ritz con la t{\'e}cnica de Monte Carlo para
evaluar integrales multidimensionales. Siendo $\Psi_T$ una funci{\'o}n
de prueba arbitraria, el principio variacional nos dice que la 
energ{\'\i}a del estado base satisface la desigualdad

\begin{eqnarray}
E_{gs} &\le& \langle\Psi_T|H|\Psi_T\rangle\nonumber\\
 &\le& \int {\rm d}^2 r_1 \dots {\rm d}^2 r_N |\Psi_T|^2 
 E_T(\vec r_1,\dots,\vec r_N),
\end{eqnarray}

\noindent
donde $H$ es el hamiltoniano del problema y $E_T=\Psi_T^{-1} H \Psi_T$.
La integral, de dimension $2 N$, es evaluada por Monte Carlo. Para ello
generamos puntos $\vec R=(\vec r_1,\dots,\vec r_N)$ con probabilidad
dada por $|\Psi_T|^2$. Tenemos:

\begin{equation}
E_{gs}\le\frac{1}{N_{puntos}}\sum_K E_T(\vec R_K).
\label{eq2}
\end{equation}

\noindent
El lado derecho de (\ref{eq2}) es el estimado para $E_{gs}$.

\begin{figure}[ht]
\vspace{0.5cm}
\begin{center}
\includegraphics[width=0.8\linewidth,angle=0]{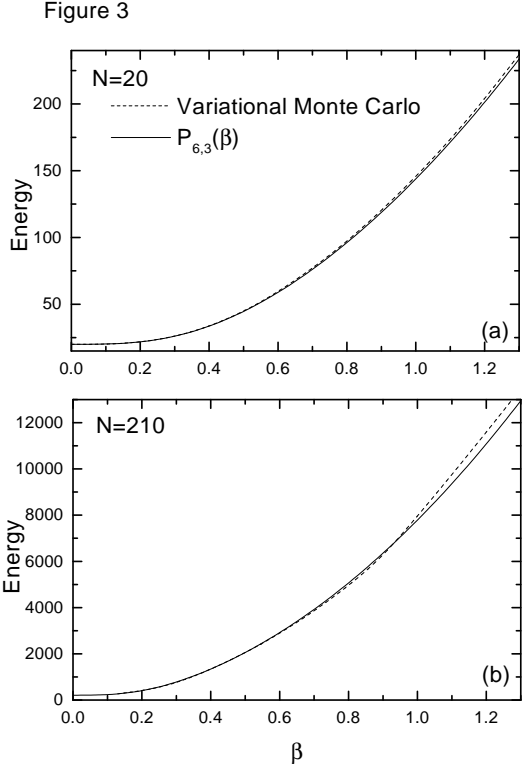}
\caption{\label{fig1} Comparaci{\'o}n entre los resultados de Monte
Carlo y los aproximantes dobles de Pad{\'e} para 20 y 210 bosones.}
\end{center}
\end{figure}

El valor de $N_{puntos}$ utilizado en nuestros c{\'alculos} fue $10^5$.
La funci{\'o}n de prueba $\Psi_T$ conten{\i}a correlaciones entre
pares de part{\'\i}culas (factores de Jastrow) y sus tres 
par{\'a}metros variacionales eran utilizados para minimizar la
energ{\'\i}a. La generaci{\'o}n de puntos en el espacio multidimensional
se hizo a partir del algoritmo de Metropolis [12]. Este es un
algoritmo secuencial en el que el punto $\vec R_{K+1}$ se genera a
partir de $\vec R_K$ de acuerdo con lo siguiente:

(i) Se crea $\vec R'=\vec R_K+\delta \vec R$, donde $\delta\vec R$ es
una perturbaci{\'o}n aleatoria.
(ii) Se genera un n{\'u}mero aleatorio $r$ distribuido uniformemente
en [0,1].
(iii) Si $|\Psi_T(\vec R')|^2/|\Psi_T(\vec R_K)|^2 > r$, entonces
$\vec R_{K+1}=\vec R'$, en caso contrario $\vec R_{K+1}=\vec R_K$.

Un ejemplo de los resultados de [1] se muestra en la Fig. \ref{fig1}.

El perfil de densidad, el potencial qu{\'\i}mico y otras magnitudes en 
sistemas de $10^4$ - $10^6$ {\'a}tomos confinados en trampas y enfriados 
por laser hasta temperaturas muy por debajo de 1K han
sido medidas experimentalmente en fecha reciente [13].

\section{Energia de sistemas multiexcitonicos a partir de la funcion BCS}

En el trabajo [2] se resalt{\'o} e hizo uso de la analog{\'\i}a entre los 
n{\'u}cleos at{\'o}micos y los sistemas de excitones en semiconductores. 
Un m{\'e}todo de c{\'a}lculo de la estructura nuclear muy empleado para 
n{\'u}cleos pesados y que se basa en una funci{\'o}n BCS [14] fue utilizado 
para calcular la energ{\'\i}a de sistemas de hasta 210 pares de electrones 
y huecos, es decir hasta 420 part{\'\i}culas, con el objetivo de investigar 
cu{\'a}ndo, o sea para qu{\'e} densidades, el sistema se comporta como un 
gas de electrones y huecos y para qu{\'e} densidades se comporta como un 
sistema de excitones descrito por la funci{\'o}n BCS. 

La funci{\'o}n de onda variacional BCS,

\begin{equation}
|BCS\rangle = \prod_{j=1}^{N_{max}} (u_j+
 v_j h_j^\dagger e_{\bar j}^\dagger)|0\rangle,
\end{equation}

\noindent
aparea huecos y electrones en estados con momento angular total
igual a cero. El estado $\bar j$ del electr{\'o}n tiene momento
invertido respecto del hueco. $|u_j|^2$ es la probabilidad de que
no exista el par $(j,\bar j)$, mientras que $|v_j|^2$ es la
probabilidad de que exista. El principio de Ritz

\begin{equation}
E_{gs} \le \langle BCS|H|BCS\rangle,
\end{equation}

\noindent
nos da un estimado para $E_{gs}$. Los coeficientes $u_j$ y $v_j$
se utilizan como par{\'a}metros variacionales sujetos a la 
ligadura $|u_j|^2+|v_j|^2=1$. La minimizaci{\'o}n conduce a la
denominada ecuaci{\'o}n del gap

\begin{equation}
\Delta_j=\sum_{k\ne j} \langle j,k|V|k,j\rangle \frac{\Delta_k}
 {2 \sqrt{\Delta_k^2+(\epsilon_k^{HF}-\mu)^2}},
\label{eq5}
\end{equation}

\noindent
donde $V$ es el potencial de interacci{\'o}n entre pares, 
$\epsilon_k^{HF}$ es la energ{\'\i}a del estado $k$ renormalizada 
por la interacci{\'o}n con el resto de los estados y $\mu$ es
el potencial qu{\'\i}mico que se determina de la ecuaci{\'o}n

\begin{equation}
N= \langle BCS|\sum_j e_j^\dagger e_j|BCS\rangle.
\end{equation}

\begin{figure}[ht]
\begin{center}
\includegraphics[width=0.7\linewidth,angle=-90]{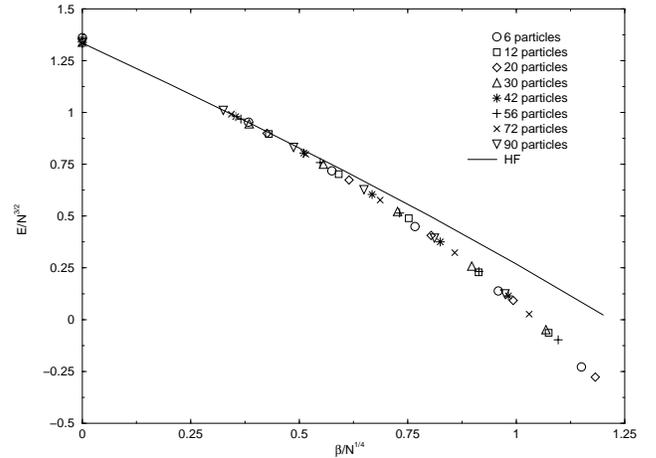}
\caption{\label{fig2} Energ{\'\i}as de sistemas multiexcit{\'o}nicos
 estimadas a partir del m{\'e}todo BCS.}
\end{center}
\end{figure}

Los coeficientes $v_j$ se han parametrizado en t{\'e}rminos 
de los $\Delta_j$ as{\'\i}:

\begin{equation}
v_j=\frac{1}{2}\left(1-\frac{\epsilon_j^{HF}-\mu}
 {\sqrt{\Delta_j^2+(\epsilon_j^{HF}-\mu)^2}}\right).
\end{equation}

Las ecuaciones (\ref{eq5}) se resuelven iterativamente y
muestran muy buenas propiedades de convergencia.

Debido a que la funci{\'o}n BCS no conserva el n{\'u}mero de 
part{\'\i}culas se hace necesario un procedimiento de proyecci{\'o}n. En 
el trabajo se emple{\'o} una proyecci{\'o}n aproximada propuesta por 
Lipkin y Nogami [15]. Al resultar evidente que esta proyecci{\'o}n era 
insuficiente se cre{\'o} un m{\'e}todo estoc{\'a}stico
basado en el algoritmo de Metropolis que ofreci{\'o} muy buenos 
resultados. En la Fig. \ref{fig2} mostramos un ejemplo de
estos resultados.

\section{Metodo estocastico para proyectar la funcion de onda BCS}

El m{\'e}todo de proyecci{\'o}n estoc{\'a}stica mencionado en la 
secci{\'o}n anterior no s{\'o}lo nos permiti{\'o} hallar mejores valores 
para la energ{\'\i}a en el caso de sistemas
de excitones, sino que de vuelta a la F{\'\i}sica Nuclear se calcul{\'o} la 
energ{\'\i}a de apareamiento en diversos modelos de estructura nuclear y 
en el is{\'o}topo de Zr con masa at{\'o}mica 100. Los resultados nuestros 
son comparables en exactitud con los mejores c{\'a}lculos cu{\'a}nticos 
por Monte Carlo de esta magnitud.

Por su valor metodol{\'o}gico, este trabajo fue publicado 
independientemente [3].

La proyecci{\'o}n de la funcion BCS sobre el sector de $N$ pares se
escribe

\begin{equation}
|\Psi_{BCS}^N\rangle = C_N \sum_{j_1,\dots,j_N}\left(
 \prod_{k=j_1}^{j_N} \frac{v_k}{u_k} a_k^\dagger 
 a_{\bar k}^\dagger\right)|0\rangle,
\end{equation}

\noindent
donde

\begin{equation}
C_N= \sum_{j_1,\dots,j_N}\left(
 \frac{v_{j_1}^2\dots v_{j_N}^2}{u_{j_1}^2\dots u_{j_N}^2} \right)^{-1/2}.
\end{equation}

\begin{figure}[ht]
\vspace{1cm}
\begin{center}
\includegraphics[width=0.65\linewidth,angle=-90]{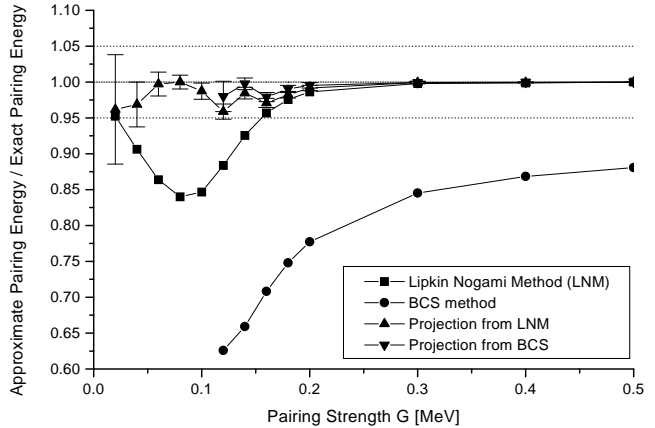}
\caption{\label{fig3} Energ{\'\i}a de apareamiento entre nucleones
 en un modelo de dos niveles degenerados.}
\end{center}
\end{figure}

El estimado para la energ{\'\i}a se obtiene de

\begin{eqnarray}
E_{BCS}^N &=& \langle \Psi_{BCS}^N|H|\Psi_{BCS}^N\rangle
 \nonumber\\
 &=&\sum_{j_1,\dots,j_N} W(j_1,\dots,j_N)\epsilon(j_1,\dots,j_N),
\label{eq10}
\end{eqnarray}

\noindent
donde $\epsilon(j_1,\dots,j_N)$ tiene interpretaci{\'o}n de 
energ{\'\i}a cuando los $N$ estados $j_1,\dots,j_N$ est{\'a}n
ocupados y los coeficientes

\begin{equation}
W(j_1,\dots,j_N)= C_N \frac{v_{j_1}^2\dots v_{j_N}^2}
 {u_{j_1}^2\dots u_{j_N}^2},
\end{equation}

\noindent
que son expl{\'\i}citamente mayores que cero, pueden
interpretarse como factores de peso. La expresi{\'o}n (\ref{eq10})
permite una evaluaci{\'o}n por Monte Carlo, donde el conjunto
de estados $(j_1,\dots,j_N)$ se generan seg{\'u}n Metropolis
con peso $W(j_1,\dots,j_N)$.

En la Fig. \ref{fig3} se muestran algunos resultados en un modelo de 
estructura nuclear.

\section{Aplicacion de la ecuacion de Bethe-Goldstone a sistemas 
de excitones}

La ecuaci{\'o}n de Bethe-Goldstone es la base de la denominada 
aproximaci{\'o}n de pares independientes en n{\'u}cleos [14]. Significa 
una mejora apreciable sobre los m{\'e}todos de Hartree-Fock o de 
part{\'\i}culas independientes en el sentido que se consideran las 
correlaciones de pares en la funci{\'o}n de onda
del n{\'u}cleo. Ha sido utilizada extensivamente 
en F{\'\i}sica Nuclear para el c{\'a}lculo de la energ{\'\i}a de 
cohesi{\'o}n de la materia 
nuclear y de la energ{\'\i}a (masa) de n{\'u}cleos intermedios.

En el trabajo [4], la ecuaci{\'o}n de Bethe-Goldstone fue reformulada para
sistemas de excitones confinados en puntos cu{\'a}nticos y empleada para el
c{\'a}lculo de la energ{\'\i}a de estos sistemas. Se investigaron sistemas 
desde el biexcit{\'o}n (2 electrones - 2 huecos) hasta conglomerados de 
12 pares, considerando ademas sistemas cargados, 4e - 2h y 6e - 2h. 
La diferencia $E_{N+1}-E_N$ entre las energ{\'\i}as de N+1 y N excitones
permite evaluar c{\'o}mo depende la frecuencia del laser que crea un nuevo
par con el n{\'u}mero de pares presentes en el punto cu{\'a}ntico. De 
nuestros resultados se dedujo adem{\'a}s la aparente inestabilidad del 
sistema 4e - 4h en un pozo cu{\'a}ntico.

\begin{figure}[ht]
\begin{center}
\includegraphics[width=0.7\linewidth,angle=-90]{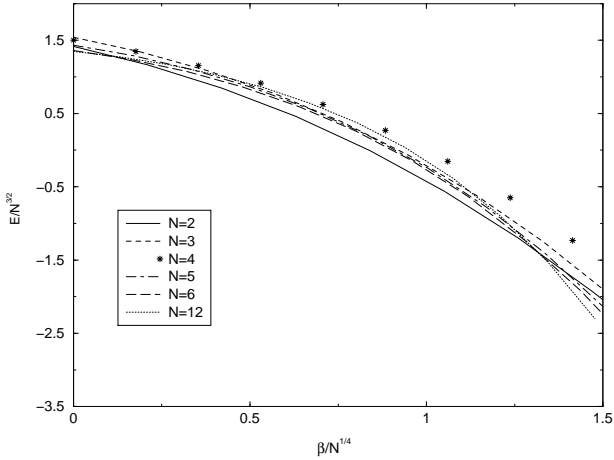}
\caption{\label{fig4} Energ{\'\i}as de sistemas de $N$ excitones
 calculadas a partir de la ecuaci{\'o}n de Bethe-Goldstone.}
\end{center}
\end{figure}

La ecuaci{\'o}n de Bethe-Goldstone describe el movimiento de un
par de fermiones que ocupan los estados $\alpha$ y $\gamma$ por
debajo del nivel de Fermi y se dispersan hacia estados por
encima del nivel de Fermi. Dicha ecuaci{\'o}n se escribe:

\begin{equation}
(T_1+T_2+Q_{\alpha\gamma}V)\psi_{\alpha\gamma}= E_{\alpha\gamma}
 \psi_{\alpha\gamma},
\end{equation}

\noindent
donde $T$ son los operadores de energ{\'\i}a de part{\'\i}culas
libres en un campo externo y $V$ es la interacci{\'o}n entre
pares de part{\'\i}culas. $Q_{\alpha\gamma}$ realiza la 
proyecci{\'o}n sobre los estados disponibles

\begin{equation}
Q_{\alpha\gamma}= |\alpha\gamma\rangle \langle \alpha\gamma|
 +\sum_{\mu',\lambda'>\mu_F} |\mu',\lambda'\rangle
 \langle\mu',\lambda'|.
\end{equation}

Escribiendo:

\begin{equation}
\psi_{\alpha\gamma}= |\alpha\gamma\rangle +\sum_{\mu',\lambda'>\mu_F} 
 C_{\mu'\lambda'}^{\alpha\gamma}|\mu',\lambda'\rangle,
\end{equation}

\noindent
obtenemos para $C_{\mu\lambda}^{\alpha\gamma}$
y $E_{\alpha\gamma}$:

\begin{eqnarray}
(\epsilon_{\mu}^{(0)}+\epsilon_{\lambda}^{(0)}-E_{\alpha\gamma})
 C_{\mu\lambda}^{\alpha\gamma}&+&\sum_{\mu',\lambda'>\mu_F} 
 \langle\mu,\lambda|V|\mu',\lambda'\rangle
 C_{\mu'\lambda'}^{\alpha\gamma}\nonumber\\
 &=&-\langle\mu,\lambda|V|\alpha,\gamma\rangle,
\label{eq15}
\end{eqnarray}

\begin{eqnarray}
E_{\alpha\gamma}&=&\epsilon_{\alpha}^{(0)}+\epsilon_{\gamma}^{(0)}
 +\langle\alpha,\gamma|V|\alpha,\gamma\rangle\nonumber\\ 
 &+&\sum_{\mu',\lambda'>\mu_F} \langle\alpha,\gamma|V|\mu',\lambda'\rangle
 C_{\mu'\lambda'}^{\alpha\gamma}.
\label{eq16}
\end{eqnarray}

Las ecuaciones (\ref{eq15}) constituyen un sistema lineal de donde
se obtiene $C_{\mu\lambda}^{\alpha\gamma}$ como funci{\'o}n de
$E_{\alpha\gamma}$, despu{\'e}s sustituyendo en (\ref{eq16})
obtenemos una ecuaci{\'o}n trascendente para $E_{\alpha\gamma}$.
La energ{\'\i}a total se halla de:

\begin{equation}
E=\sum_{\alpha}\epsilon_{\alpha}^{(0)}
 +\sum_{\alpha<\gamma} (E_{\alpha\gamma}-
 \epsilon_{\alpha}^{(0)}-\epsilon_{\gamma}^{(0)}).
\end{equation}

En la Fig. \ref{fig4} mostramos la energ{\'\i}a de $N$ excitones 
$(2\le N\le 12)$ calculadas por este m{\'e}todo.

\section{Absorcion y luminiscencia inter-banda en puntos cuanticos de 
electrones}

En el trabajo [5] se calcul{\'o} el coeficiente de absorci{\'o}n, la 
luminiscencia y el perfil de densidades de electrones y huecos en puntos 
cu{\'a}nticos peque{\~n}os (de 0 a 3 electrones)
en presencia de campos magn{\'e}ticos muy intensos, entre 8 y 50 Teslas.  
Experimentos de luminiscencia en pozos cu{\'a}nticos (donde existen 
infinitos electrones confinados en una regi{\'o}n cuasi-bidimensional) 
bajo campos magn{\'e}ticos tan altos han sido reportados recientemente [16]. 

Los estado finales de un proceso de absorci{\'o}n e iniciales en uno de 
luminiscencia contienen un par electr{\'o}n - hueco adicional. El sistema
mas grande considerado fue, por tanto, el 4e - 1h. Las funciones de
onda de onda de los estados iniciales y finales fueron obtenidas por
diagonalizaci{\'o}n exacta de la ecuaci{\'o}n de Schrodinger en bases de 
aproximadamente 5,000 funciones.

Se obtuvieron resultados interesantes como la re-estructuraci{\'o}n de los
estados de mas baja energ{\'\i}a al variar el campo magn{\'e}tico, dando 
como resultado el corrimiento de los picos de absorci{\'o}n y de 
luminiscencia, as{\'\i} como variaciones abruptas del coeficiente de 
absorci{\'o}n o de la luminiscencia para valores de campos 
correspondientes a los llenados 1/3, 1/5, etc del efecto Hall cu{\'a}ntico.

\begin{figure}[ht]
\vspace{1cm}
\begin{center}
\includegraphics[width=0.7\linewidth,angle=-90]{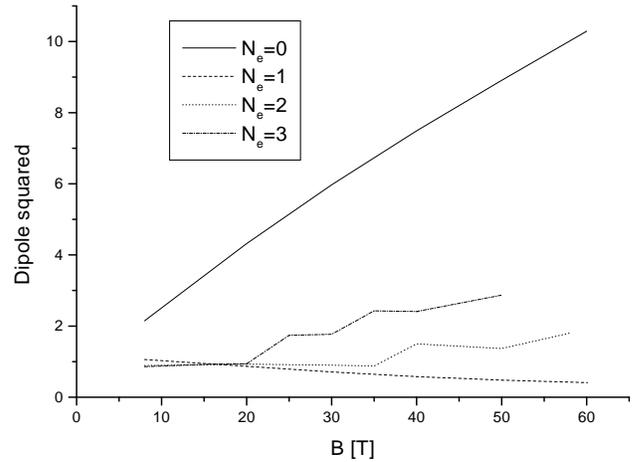}
\caption{\label{fig5} Intensidad del pico principal de luminiscencia 
 correspondiente a excitones m{\'u}ltiplemente cargados.}
\end{center}
\end{figure}

En la Fig. \ref{fig5} mostramos los m{\'a}ximos de luminiscencia
correspondientes a multiexcitones con un exceso de carga $N_e$, es
decir la recombinaci{\'o}n desde sistemas con $N_e+1$ electrones
y un hueco.

\section{Resonancias dipolares gigantes en puntos cuanticos}

Las resonancias dipolares gigantes (GDR) son excitaciones colectivas de 
n{\'u}cleos [17] que juegan un importante papel en la foto-absorci{\'o}n y 
en otras reacciones nucleares. En el trabajo [6] se muestra que estados
muy similares existen en sistemas de electrones y huecos en puntos
cu{\'a}nticos semiconductores.

Las energ{\'\i}as de excitaci{\'o}n y los elementos de matriz del operador 
de dipolo son calculados a partir de la denominada aproximaci{\'o}n de 
fase aleatoria (RPA) en sistemas de hasta 110 pares e - h, 
concluy{\'e}ndose que en las GDR se concentra aproximadamente el 98 \% 
de la absorci{\'o}n de luz infra-roja en estos sistemas. 

La RPA requiere como paso previo que el estado base
sea obtenido en la aproximaci{\'o}n de Hartree - Fock (HF), por lo que
se implement{\'o} la solucion iterativa de las ecuaciones de HF

\begin{eqnarray}
&\sum_t &\left\{ E_{es}^{(0)}\delta_{st} + \beta\sum_{\gamma\le \mu_F^e} 
 \sum_{u,v} \left[ \langle s,u|1/r|t,v\rangle\right.\right.\nonumber\\
 &-&\left.\langle s,u|1/r|v,t\rangle
 \right]C_{\gamma,u}^e C_{\gamma,v}^e \label{HF}\\
 &-& \left. \beta \sum_{\gamma\le\mu_F^h}\sum_{u,v} 
 \langle s,u|1/r|t,v\rangle C_{\gamma,u}^h C_{\gamma,v}^h\right\}
 C_{\alpha,t}^e=E_{e \alpha} C_{\alpha,s}^e\nonumber,
\end{eqnarray}

\noindent
donde los orbitales electr{\'o}nicos ocupados se expresan en t{\'e}rminos de las 
funciones de oscilador a trav{\'e}s de los coeficientes $C_{\alpha,s}$ . Ecuaciones 
an{\'a}logas a (\ref{HF}) se pueden escribir para la descomposici{\'o}n de los
orbitales correspondientes a los huecos. Note que los $E^{(0)}$ son las energ{\'\i}as
del oscilador en 2D

\begin{eqnarray}
E_{es}^{(0)}&=&\hbar\sqrt{\omega_0^2+\omega_c^2/4}~\{ 2 k_s+|l_s|+1\}+
 \frac{\hbar\omega_c}{2} l_s\nonumber\\
 &&+g_e\mu_B B S_z^e,\\
E_{hs}^{(0)}&=&\frac{m_e}{m_h}~\hbar\sqrt{\omega_0^2+\omega_c^2/4}~
 \{ 2 k_s+|l_s|+1\}\nonumber\\
 &&-\frac{m_e}{m_h}\frac{\hbar\omega_c}{2} l_s-g_h\mu_B B S_z^h,
\end{eqnarray}

\begin{figure}[ht]
\vspace{0.5cm}
\begin{center}
\includegraphics[width=0.7\linewidth,angle=-90]{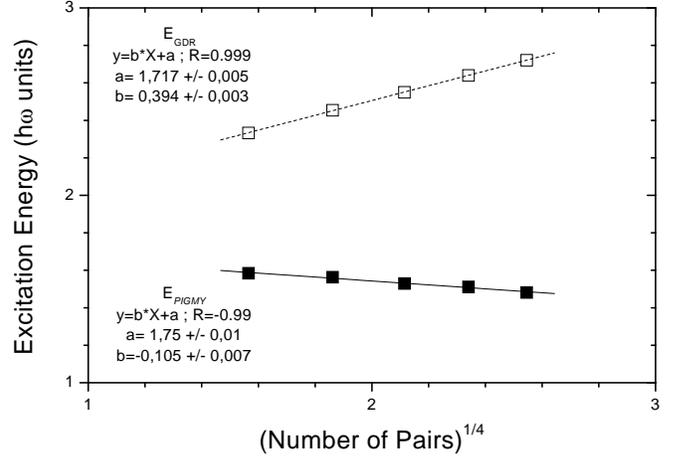}
\caption{\label{fig6} Posici{\'o}n de las resonancias gigantes y pigmeas 
 como funci{\'o}n del n{\'u}mero de pares en el punto cu{\'a}ntico.}
\end{center}
\end{figure}

Por otro lado, en la RPA el estado base $|RPA\rangle$ se supone que contiene
correlaciones entre las part{\'\i}culas y los estados excitados se buscan en la
forma

\begin{equation}
\Psi = Q^{\dagger} |RPA\rangle ,
\end{equation}

\noindent
donde el operador $Q^{\dagger}$  viene dado por la expresi{\'o}n

\begin{equation}
Q^{\dagger}=\sum_{\sigma,\lambda}(X^e_{\sigma\lambda} 
 e^{\dagger}_{\sigma} e_{\lambda}+X^h_{\sigma\lambda}
 h^{\dagger}_{\sigma} h_{\lambda}-Y^e_{\lambda\sigma} 
 e^{\dagger}_{\lambda} e_{\sigma}-Y^h_{\lambda\sigma} 
 h^{\dagger}_{\lambda} h_{\sigma}).
\label{Q}
\end{equation}

El {\'\i}ndice $\lambda$ corre sobre estados ocupados de HF, 
mientras que $\sigma$ corre sobre estados desocupados (por encima del
nivel de Fermi). Los coeficientes $X,~Y$ se determinan por ecuaciones
del tipo:

\begin{eqnarray}
\sum_{\tau,\mu}\left\{ A^{ee}_{\sigma\lambda,\tau\mu} X^e_{\tau\mu}\right.
 &+&A^{eh}_{\sigma\lambda,\tau\mu} X^h_{\tau\mu}
 +B^{ee}_{\sigma\lambda,\mu\tau} Y^e_{\mu\tau}\nonumber\\
 &+&\left. B^{eh}_{\sigma\lambda,\mu\tau} Y^h_{\mu\tau}\right\}=
 \hbar\Omega X^e_{\sigma\lambda},
\label{RPAeq}
\end{eqnarray}

\noindent
en las cuales $\hbar\Omega$ son las energ{\' \i}as de excitaci{\'o}n y las matrices
$A$ y $B$ vienen dadas por

\begin{eqnarray}
A^{ee}_{\sigma\lambda,\tau\mu} &=& (E_{e\sigma}-
 E_{e\lambda}) \delta_{\sigma\tau} \delta_{\lambda\mu}
 + \beta \left( \langle \sigma,\mu |1/r|\lambda,\tau\rangle \right.\nonumber\\
 &&-\langle \left.\sigma,\mu |1/r|\tau,\lambda\rangle \right),\nonumber\\
A^{eh}_{\sigma\lambda,\tau\mu} &=& -\beta\langle \sigma,\mu |
 1/r|\lambda,\tau \rangle,\nonumber\\
B^{ee}_{\sigma\lambda,\mu\tau} &=& \beta ( \langle \sigma,\tau
|1/r|\lambda,\mu\rangle -\langle \sigma,\tau |1/r|
\mu,\lambda\rangle ),\nonumber\\
B^{eh}_{\sigma\lambda,\mu\tau} &=& -\beta\langle \sigma,\tau 
|1/r|\lambda,\mu \rangle. 
\end{eqnarray}

Usualmente, energ{\'\i}as de excitaci{\'o}n positivas (f{\'\i}sicas) y negativas (no f{\'\i}sicas) 
aparecen como soluciones de (\ref{RPAeq}). Las soluciones f{\'\i}sicas aniquilan
el estado base 

\begin{equation}
Q |RPA\rangle = 0,
\end{equation}

\noindent
y satisfacen la condici{\'o}n de normalizaci{\'o}n 

\begin{equation}
1= \sum_{\sigma,\lambda} 
 \{ |X_{\sigma\lambda}^e|^2+|X_{\sigma\lambda}^h|^2
 -|Y_{\lambda\sigma}^e|^2-|Y_{\lambda\sigma}^h|^2 \} .
\end{equation}

Las ecuaciones (\ref{RPAeq}) dan directamente las energ{\'\i}as
de excitaci{\'o}n de los estados excitados. Los elementos de
matriz del operador dipolar, que causa la transici{\'o}n del
estado base a los estados excitados, se expresan en t{\'e}rminos de
los coeficientes $X$ e $Y$. La posici{\'o}n de la GDR, es decir del estado
que absorbe casi toda la fortaleza de la transici{\'o}n,  como funci{\'o}n del 
n{\'u}mero de pares electr{\'o}n-hueco se muestra en la Fig. \ref{fig6}.

\section{Las denominadas resonancias pigmeas}

En fecha reciente, se ha observado experimentalmente [18], que el 2 \%
de la foto-absorci{\'o}n en n{\'u}cleos se concentra en estados de mas 
baja energ{\'\i}a de excitaci{\'o}n. Por contraposici{\'o}n a las GDR, 
a estos estados se les ha llamado ``resonancias pigmeas''. 

\begin{figure}[ht]
\begin{center}
\includegraphics[width=0.8\linewidth,angle=0]{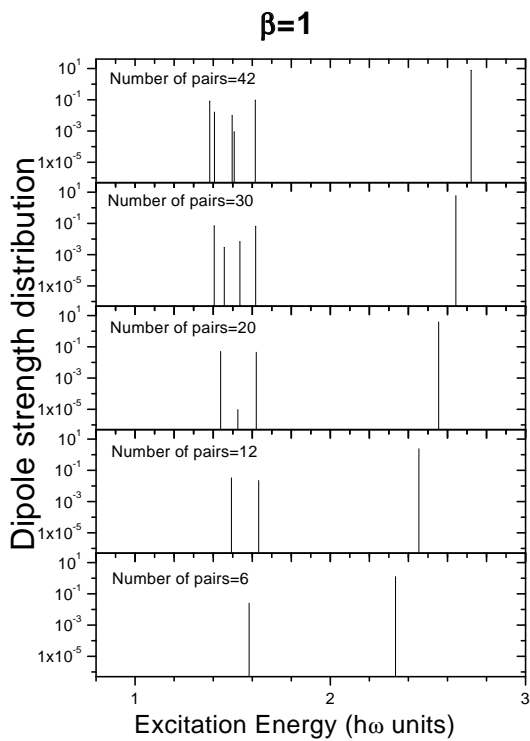}
\caption{\label{fig7} Intensidades de las resonancias gigantes y 
 pigmeas en puntos cu{\'a}nticos.}
\end{center}
\end{figure}

Los resultados del trabajo [7] muestran que la analog{\'\i}a con 
n{\'u}cleos es casi perfecta pues los c{\'a}lculos de la absorci{\'o}n 
en sistemas de electrones y huecos tambi{\'e}n dan como resultado una 
concentraci{\'o}n del 2 \% restante en estados de baja energ{\'\i}a de 
excitaci{\'o}n. (Ver Fig. \ref{fig7}).

El n{\'u}mero de resonancias pigmeas, al igual que en n{\'u}cleos, 
aumenta con el n{\'u}mero total de part{\'\i}culas. La posici{\'o}n de la 
GDR y del promedio de las resonancias pigmeas tambi{\'e}n depende del 
n{\'u}mero N de pares e - h en el punto cu{\'a}ntico. Esta dependencia es 
cualitativamente similar a la de n{\'u}cleos, es decir la GDR se desplaza 
hacia energ{\'\i}as mas altas al aumentar N, mientras
que las resonancias pigmeas se desplazan hacia mas bajas energ{\'\i}as.

\section{Tunelamiento resonante a traves de un punto cuantico}

Experimentos recientes han mostrado la factibilidad de controlar el 
n{\'u}mero de electrones en un punto cu{\'a}ntico (desde uno hasta 
cuarenta), a la vez que reportan mediciones de la conductancia en los 
mismos [19].

En el trabajo [8] se tomaron par{\'a}metros que reprodujeran 
aproximadamente las condiciones de los experimentos en [19] y se 
calcul{\'o} exactamente la 
conductancia y la densidad de niveles para un punto de 6 electrones. Hay
que resaltar que el punto cu{\'a}ntico se halla separado de los electrodos 
por barreras de potencial, por lo que la conducci{\'o}n a trav{\'e}s del 
mismo se realiza por medio de un proceso de tunelamiento resonante. 

Las energ{\'\i}as y funciones de onda de 6 y 7 electrones en el punto 
fueron obtenidas a partir de la diagonalizaci{\'o}n de la ecuaci{\'o}n 
de Schrodinger en bases de hasta 40,000 funciones utilizando el m{\'e}todo 
de Lanczos para tratar matrices tan grandes.

\begin{figure}[ht]
\begin{center}
\includegraphics[width=0.8\linewidth,angle=0]{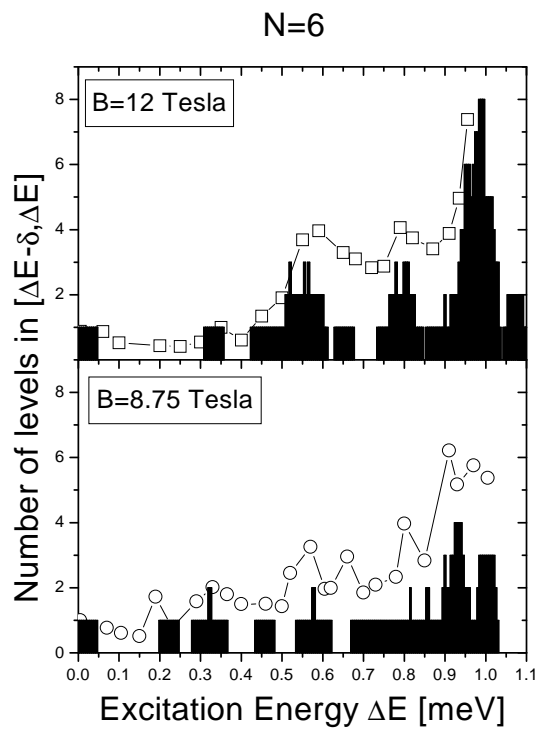}
\caption{\label{fig8} Densidad de niveles de energ{\'\i}a para un punto
 cu{\'a}ntico con 6 electrones. Las barras representan los resultados de
 la diagonalizaci{\'o}n exacta del hamiltoniano, mientras que los 
 s{\'\i}mbolos son estimados a partir del coeficiente de trasmisi{\'o}n.}
\end{center}
\end{figure}

Nuestros c{\'a}lculos revelan que a partir de una medici{\'o}n precisa de 
la conductancia se puede obtener la densidad de estados excitados en 
sistemas de pocos electrones. Esta afirmaci{\'o}n se ilustra en la Fig.
\ref{fig8}, donde la densidad de estados obtenida por diagonalizaci{\'o}n 
exacta es comparada con la que se obtiene a partir de la conductancia
(coeficiente de trasmisi{\'o}n) calculada para un punto con 6 electrones.

El m{\'e}todo de Lanczos hace uso del algoritmo de Graham-Schmidt para
ortogonalizar una base de vectores que es creada por la propia matriz
hamiltoniana (que se supone sim{\'e}trica). Debido a que la base es creada 
por la propia matriz, las
propiedades de convergencia ante truncamiento son muy buenas. Quiere decir
que para obtener los primeros 30 autovalores con buena aproximaci{\'o}n 
basta representar la matriz en el subespacio de los primeros 500 vectores 
de Lanczos, aun cuando la dimensi{\'o}n de la matriz sea 40,000.

Sea $\vec e_1$ un vector arbitrario. La 1ra iteraci{\'o}n de Lanczos consiste
en lo siguiente:

Calculamos $a_1=\vec e_1\cdot H \vec e_1$, $\vec v_2=H \vec e_1-a_1 \vec e_1$,
$b_2=|\vec v_2|$ y definimos $\vec e_2=\vec v_2/b_2$. Tenemos entonces:

\begin{equation}
H \vec e_1=a_1 \vec e_1+b_2 \vec e_2.
\end{equation}

En la 2da iteraci{\'o}n calculamos: $a_2=\vec e_2\cdot H \vec e_2$, 
$\vec v_3=H \vec e_2-a_2 \vec e_2-b_2 \vec e_1$,
$b_3=|\vec v_3|$ y definimos $\vec e_3=\vec v_3/b_3$. Tenemos:

\begin{equation}
H \vec e_2=b_2 \vec e_1+a_2 \vec e_2+b_3 \vec e_3.
\end{equation} 

Las siguientes iteraciones se hacen de manera an{\'a}loga.
Notice que en la base $\{\vec e_i\}$ el operador $H$ se 
representa por una matriz tridiagonal.

\section{Polarizacion del espin y luminiscencia en sistemas de excitones}

En el trabajo [9] se analizaron los efectos combinados de la energ{\'\i}a 
de confinamiento del punto cu{\'a}ntico, la energ{\'\i}a Zeeman, la 
interacci{\'o}n de Coulomb y de campos magn{\'e}ticos intensos sobre la 
polarizaci{\'o}n de los espines electr{\'o}nicos y la luminiscencia 
coherente en sistemas de hasta 40 pares e - h.

\begin{figure}[h]
\begin{center}
\includegraphics[width=0.8\linewidth,angle=0]{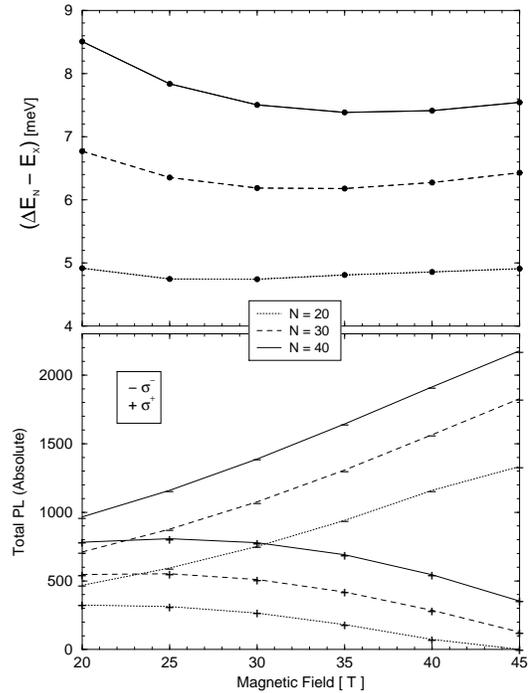}
\caption{\label{fig9}
 Posici{\'o}n e intensidad de la l{\'\i}nea de luminiscencia como
 funci{\'o}n del campo magn{\'e}tico y el n{\'u}mero de excitones en el 
 punto. $E_X$ es la energ{\'\i}a del excit{\'o}n.}
\end{center}
\end{figure}

El esquema te{\'o}rico ya hab{\'\i}a sido utilizado en [3], es decir una 
funci{\'o}n de onda BCS con proyecci{\'o}n aproximada a lo Lipkin - Nogami. 
La bondad de este m{\'e}todo en el caso de campos magn{\'e}ticos grandes 
consiste en que la funci{\'o}n BCS reproduce el estado b{\'a}sico exacto -- 
un condensado de excitones [20]-- en el l{\'\i}mite de campos infinitos. 

Mediciones de la luminiscencia en pozos cu{\'a}nticos como funci{\'o}n de 
la intensidad y polarizaci{\'o}n de la luz que crea
los excitones hab{\'\i}an sido reportadas en [21]. 

Nuestros c{\'a}lculos arrojan s{\'o}lo un 10 \% de polarizaci{\'o}n neta 
de los espines electr{\'o}nicos, aun a campos tan altos como 20 Teslas, 
debido a la fuerte interacci{\'o}n entre electrones y huecos. Adem{\'a}s, 
se predice un corrimiento hacia el azul del pico de luminiscencia 
coherente con el aumento del n{\'u}mero de pares e - h presentes en el 
punto (o lo que es equivalente, aumento de la intensidad del laser).

\section{Transiciones opticas intra-bandas en el biexciton}

Las transiciones entre estados electr{\'o}nicos internos, causadas por
absorci{\'o}n de fotones en el infra-rojo (5 - 10 meV), han sido 
estudiadas recientemente por medio de una t{\'e}cnica, conocida como ODR, 
consistente en observar sus efectos en las l{\'\i}neas de luminiscencia 
[22]. Los experimentos han sido realizados en pozos cu{\'a}nticos y la 
l{\'\i}nea de luminiscencia tomada como referencia es la l{\'\i}nea 
fundamental del excit{\'o}n.

En el trabajo [10], se calculan los niveles de energ{\'\i}a, funciones de 
onda y elementos matriciales del operador dipolar para el biexcit{\'o}n. 
Campos magneticos entre 0 y 5 Teslas son considerados. Los par{\'a}metros
utilizados (masas de las part{\'\i}culas, etc) corresponden a los de los
experimentos [22]. La mezcla de sub-bandas de valencia se consider{\'o}
aproximadamente. El m{\'e}todo de c{\'a}lculo fue la diagonalizaci{\'o}n 
directa de la ecuaci{\'o}n de Schrodinger en bases de hasta 
10,000 funciones.

\begin{figure*}[ht]
\begin{center}
\includegraphics[width=0.6\linewidth,angle=-90]{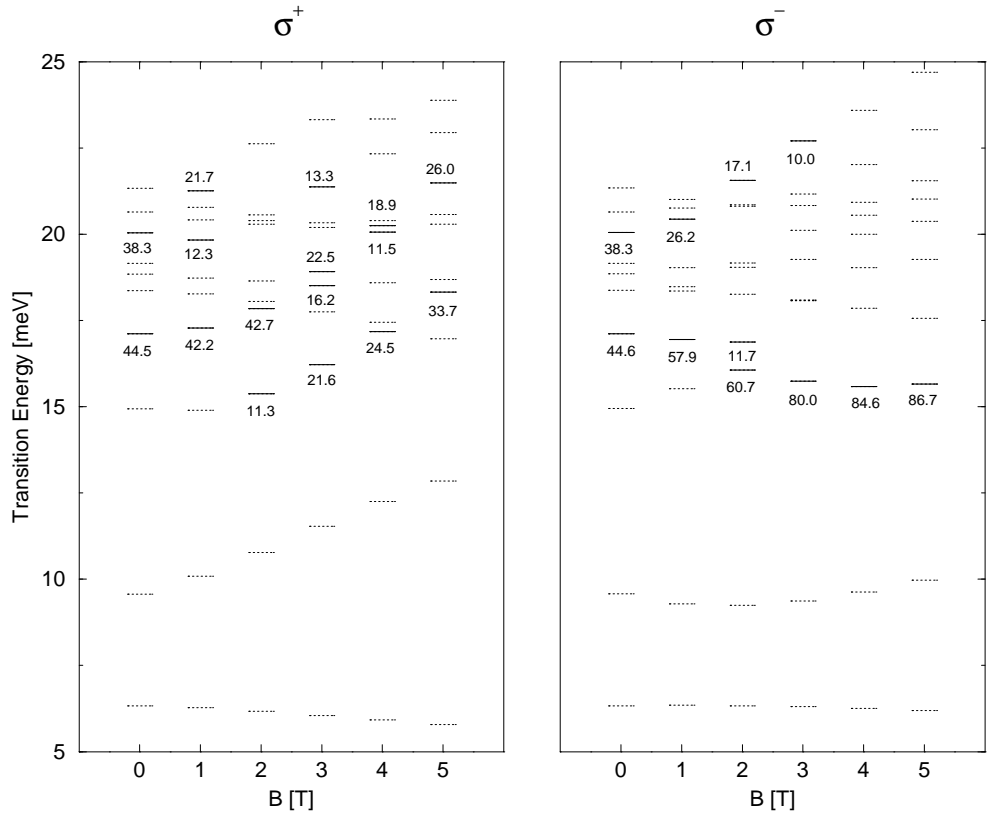}
\caption{\label{fig10}
 Energ{\'\i}as de los estados de llegada y probabilidades 
 relativas de absorci{\'o}n (s{\'o}lo las superiores al 10 \% se 
 representan) como funci{\'o}n del campo magn{\'e}tico.}
\end{center}
\end{figure*}

Se hicieron predicciones para las posiciones e intensidades de las
transiciones internas en el biexcit{\'o}n, las cuales deben ser observadas 
siguiendo la linea biexcit{\'o}nica en la ODR. Se obtuvieron resultados 
interesantes sobre anti-cruzamiento de niveles y transferencia de
fuerza de oscilador entre estados que colisionan. Un ejemplo de estos
resultados se muestra en la Fig. \ref{fig10}.

\section{Conclusiones}

Se han presentado resultados sobre propiedades f{\'\i}sicas de sistemas
confinados del estado s{\'o}lido en que intervienen decenas o cientos de
part{\'\i}culas. Se ha puesto de manifiesto la gran analog{\'\i}a de 
estos sistemas con los n{\'u}cleos at{\'o}micos, lo cual fundamenta la 
aplicaci{\'o}n de m{\'e}todos de la F{\'\i}sica Nuclear en la 
investigaci{\'o}n de los mismos. 

\section{Agradecimientos}

Los resultados reportados corresponden a trabajos conjuntos con
R. Capote, R. P{\'e}rez, B. Rodr{\'\i}guez, A. Delgado, E. 
Men{\'e}ndez, B. Partoens, A. Matulis, F.M. Peeters, L.
Quiroga y F.J. Rodr{\'\i}guez.

\end{document}